\documentclass{article} 
\usepackage{iclr2019_conference,times}
\usepackage[utf8]{inputenc}

\usepackage{graphicx}
\usepackage{dcolumn}
\usepackage{bm}
\usepackage[mathlines]{lineno}
\usepackage{xcolor}
\usepackage{hyperref}
\usepackage{url}
\usepackage{breakurl}
\usepackage{graphicx}
\usepackage{caption}
\usepackage{amsmath}
\usepackage{cleveref}
\usepackage{textgreek}
\usepackage{xspace}
\usepackage[binary-units]{siunitx}
\usepackage{amsmath}
\usepackage{booktabs}
\usepackage{listings}
\usepackage{xcolor}
\usepackage{verbatim}
\usepackage{subfigure}
 
\title{Modeling vehicular mobility patterns using recurrent neural networks}

\iclrfinalcopy

\author{Kevin O'Keeffe \\
Senseable City Lab, Massachusetts Institute of Technology \\
Cambridge, MA 02139 \\
\texttt{kokeeffe@mit.edu} 
\And
Paolo Santi \\
Senseable City Lab, Massachusetts Institute of Technology \\
Cambridge, MA 02139 \\
\And 
Carlo Ratti \\
Senseable City Lab, Massachusetts Institute of Technology \\
Cambridge, MA 02139 \\
}

\date{\today}

\begin{document}
\maketitle



\begin{abstract}
Data on vehicular mobility patterns have proved useful in many contexts. Yet generative models which accurately reproduce these mobility patterns are scarce. Here, we explore if recurrent neural networks can cure this scarcity. By training networks on taxi from NYC and Shanghai, and personal cars from Michigan, we show most aspects of the mobility patterns can be reproduced. In particular, the spatial distributions of the street segments usage is well captured by the recurrent neural networks, which other models struggle to do.
\end{abstract}

\section*{Introduction}

Data on vehicular mobility patterns has inspired much academic research. Large datasets on taxis, for example, have allowed researchers to analyze the benefits of ride and vehicle sharing \cite{santi2014quantifying,barann2017open,yu2017environmental,tachet2017scaling,vazifeh2018addressing,wallar2019optimizing} and to quantify the feasibility of taxi-based mobile sensing \cite{o2019quantifying}. These data have also been used to train a wide range of machine learning models such as those which predict passenger demand \cite{yao2018deep, rodrigues2019combining, liu2019contextualized}, passenger destination \cite{lv2018t,zhang2018multi}, traffic density \cite{castro2012urban,niu2014deepsense}, and human mobility \cite{li2012prediction,tang2015uncovering,zheng2016two}. And data on personal cars has, for example, led to driver classification methods useful for risk profiling and accident prediction \cite{fugiglando2018driving,massaro2016car,martinelli2018human} .

Yet research on generative models of vehicular mobility patterns -- models which produce synthetic data that mimics real-world data  -- is less developed. Such generative models could have great applied utility, potentially allowing researchers to design realistic simulators for A/B testing (which could be useful to policy makers), or for use in taxis-related research in cities for which data are currently unavailable.

In this work, we explore if recurrent neural networks (RNN) can be used to generate realistic mobility patterns for taxis and personal cars. Following \cite{wu2017modeling}, we map vehicles motion onto a street network so that trajectories can be represented as sequences of nodes. Posed in this manner, generating the trajectories becomes equivalent to a `sequence-to-sequence' prediction problem -- problems which RNN`s are specifically designed to solve. In \cite{wu2017modeling}, RNNs (trained on data from taxis only) were shown to have good classification accuracy. But aside from classification accuracy no other properties of taxis' movements were studied. Here, we extend their work by analyzing the full, global, mobility patterns of fleets of both taxis and personal cars whose movements are modeled by RNN's. Specifically, we study the distributions of \textit{street popularities} $p_i$, defined as the relative number of times the $i$-th street (or node) is traversed by a fleet of taxis moving on the street network during a given reference period. These $p_i$ are key quantities for certain use cases \cite{o2019quantifying}, and are an accurate way to characterize taxis' mobility patterns. 


\section*{Datasets}
Our results are based on three datasets: taxi data from New York City (confined to the borough of Manhattan), taxi data from Shanghai, and personal car data from Michigan. The data from NYC are incomplete; only the origin and destination of each trip were known, so we used a method to generate the full trajectory which used hour by hour estimates of traffic congestion (see SM). The Shanghai / Michigan data however contained the full trace of a taxi / personal car as it serves its passenger / drives around (note for taxi datasets data is only available when passengers are on board only. The passenger-seeking part of a taxis movements is unavailable). Given the large size of Shanghai, we split the dataset into six sub-cities within Shanghai: Yangpu, Changning, Hongkou, Jingan, Putuo, Xuhui (Note Shanghai contains other sub-cities; these six were chosen arbitrarily) which were easier to work with computationally. In all datasets the vehicles positions are given as GPS coordinates, so we snapped the GPS coordinates to the nearest node of the given street network using the python package OSMNX \cite{boeing2017osmnx}, which was also used to define the street network itself. This way, a trajectory $T$ was represented as a sequence of nodes, $T = (S_{1}, S_{2}, \dots)$, where $S_i \in \mathcal{S} := (1,2,\dots, N_S)$ label segments and there are $N_S$ segments in total, a format suitable for RNN's. We further describe the datasets in the SM. 

\section*{Results}
Our goal is to generate trajectories $T_j = (S_{j_1}, S_{j_2}, \dots, S_{j_L})$ of arbitrary length $L$. As in standard in sequence-to-sequence problems,  we formulate this as a series of multi-class classification problems. First, we predict the $i$-th segment in a sequence, $y = S_i \in \mathcal{S}$, given the $k$ previous segments $\mathbf{x} = (S_{i-1},S_{i-2},\dots,S_{i-k})$. This is done as $y = argmax(\mathbf{z})$, where $\mathbf{z}$ are the class probabilities outputed by the model $\mathbf{z} = f(\mathbf{x})$, where $f$ represents the RNN. Then we update the input sequence to $\mathbf{x} = (S_{i}, \dots, S_{i-k-1})$ and predict $S_{i+1}$, and then repeat until $L$ new segments have been predicted / generated. The input sequence $\mathbf{x}$ can be optionally endowed with other features before being fed into the RNN. We chose the GPS coordinates of a given segment and the hour at which the sequence was traversed. We assumed that all segments $S_i$ in a given sequence $\mathbf{x}$ were traversed at the same hour for convenience. This seemed a reasonable approximation since as shown in Figure~\ref{fig:trip-length} in Appendix~\ref{appendix:sf}, the mean and median trip lengths are $10$ and $12$ mins respectively (also notice most trips are $\leq$ 30 mins). Of course even if a trip is short in duration, if it begins near the end of an hour, the approximation will break down. So we ran experiments both with and without using time as a feature and found no significant difference in the final performance.

Notice that in the above formulation the predicted segments can be any of the $N_{S}$ street segments $y = S_i \in \mathcal{S} = (1,\dots, N_S)$. This is a weakness of the formulation because in reality segments are contiguous in space, meaning that a given segment a $S_i$, only its neighbours $N_{S_i}$ should be predictable. In \cite{wu2017modeling}, a variant of an RNN called `CSSRNN' (constrained state space RNN) was introduced which imposes these constraints. However in our experiments, we found that a simple RNN effectively learns to avoid unphysical predictions, so we used this simpler model throughout our paper (Moreover, the CSSRNN improves the accuracy of the regular RNN by just $ \leq \% 1 $ as show in Table 2 in \cite{wu2017modeling}).

\begin{figure*}[t!]
    \centering
    \includegraphics[width=0.85 \linewidth]{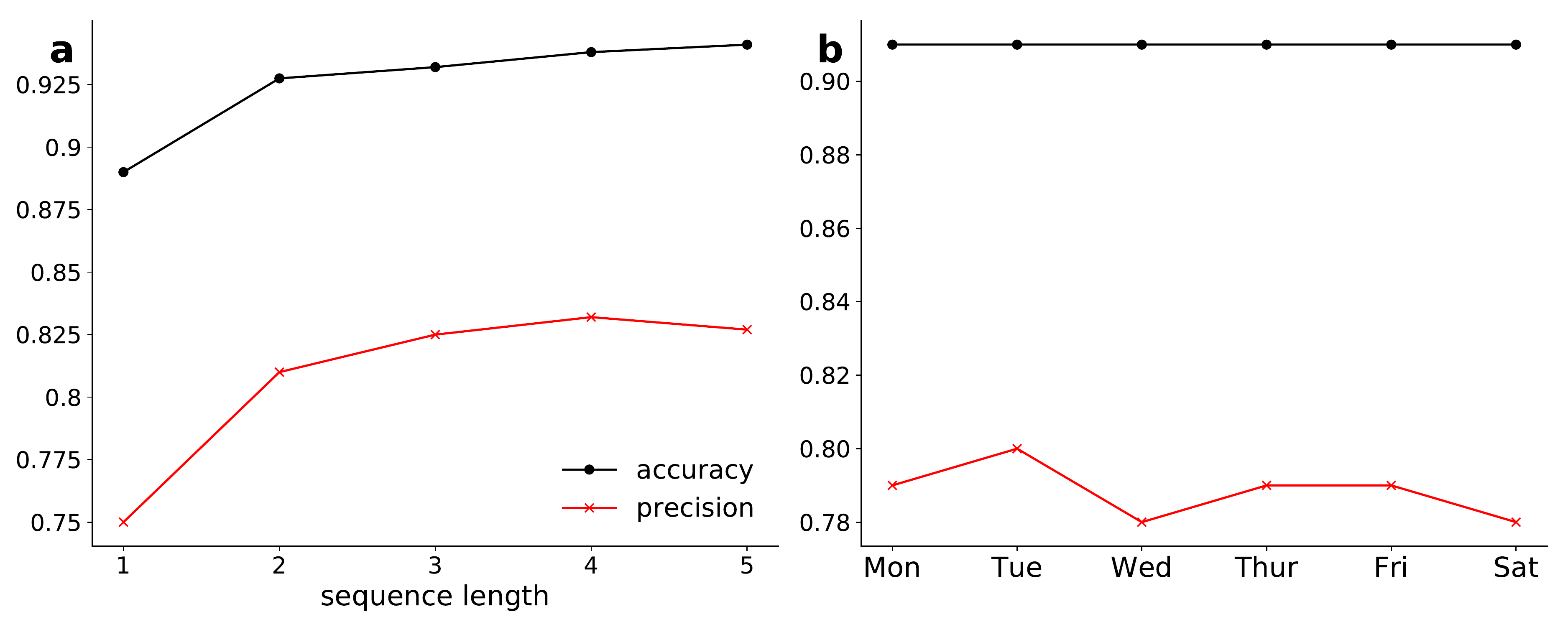}
    \caption{(a) Accuracy and precision versus sequence length for NYC data (see Table~\ref{tab:classification} for a description of the data). As can be seen, there is little increase in accuracy and precision beyond a sequence length of $3$ so we used this value throughout our work. (b) Accuracy and precision versus trips taken from different days of week starting 01/18/11; no significant variation is present. The simulation parameters used for both panels are the same as those described in the Results section.}
    \label{fig:seqLength}
\end{figure*}

As described above, our datasets consist of stacks of trajectories. We chopped these into sequences of length $k+1$, the first $k$ of which comprised the input data $\mathbf{x}$ and final element being the outcome $y$. We then divided the data into a 90-10 train-test split. A further $10 \%$ of the training data was used to validate hyperparameters. The model architecture we used was: 1 embedding layer, 2 LSTM layers, 1 Dense layer with ReLu activation, and 1 Dense layer with softmax activation. This architecture is commonly used in other sequence-to-sequence problems such as natural language modeling. The embedding layer maps the input segments (represented as integers) into a vector space of given dimension, on which the LSTMS most naturally operate. The dense layer then maps the output from the LSTM's into class probabilities. We used both unidirectional (i.e. regular) and bidirectional LSTM's (which are easily implemeted in the python tensorflow wrapper \textit{keras}). We used an embedding dimension of $256$ and hidden dimension of $64$. An Adam optimizer was used with learning rate $1e-4$ for $40$ epochs with batch size = $64$. As discussed in Appendix~\ref{appendix:ho}, all these hyperparameters were found by optimizing the validation accuracy. Finally, we choose a sequence length of $k = 3$, since, as shown in Figure~\ref{fig:seqLength}(a), the marginal gains in performance were small beyond this value. 


We use the following metrics to the assess the performance of our model: accuracy, precision, perplexity, and a home-made parameter $\gamma_{n}$. Formal definitions are given in Appendix~\ref{appendix:metrics}. Accuracy and precision are standard choices and give general information about the quality of the model. But they are limited in the sense that a low accuracy does not necessarily mean poor performance in the context of trajectory generation. This is because a given $\mathbf{x}$ does not map to a single $y$ since the $i$-th segment $S_i$ in the sequence $(S_{i-k}, \dots, S_{i-1})$ is not unique (such as at an intersection). So we also computed the perplexity over the test data, as is common in sequence generation problems (i.e. natural language generation). $\gamma_n$ measures how well the model avoids unphysical predictions (as discussed above). It is the average fraction of illegal predictions in the top $n$ predictions. So $\gamma_1 = 1$ means the most probable $y_i$ is a physically possible street, and $\gamma_3 = 3$ means all of the top three predictions are physically possible streets.

Table~\ref{tab:classification} shows the RNN has good performance, with high accuracies, high precisions, and low perplexities achieved on most datasets. The metrics for NYC are better than those from the Shanghai sub-cities. This makes sense, real-world trajectories presumably being harder to predict that synthetic (recall the trajectories from NYC were generated  whereas those from Shanghai all come from data). The Michigan data has lowest performance (in particular has very high preplexity of 14.4.) This is perhaps due to the smaller size of the dataset ($O(10^4)$ sequences as compared to $O(10^5)$ for the other cities; see caption in Table~\ref{tab:classification}). Interestingly, the results when GPS coordinates and time of day were used as additional features do not improve the results. (For greater clarity we only report results with additional features and on for NYC, but the results from other cities showed the same trend). We also checked if the results showed any temporal variations by computing the accuracy and precision on datasets from different days of the week in NYC (as mentioned in the caption of Table~\ref{tab:classification} the metrics from NYC are based on data from a single day.) Figure~\ref{fig:seqLength}(b) shows they do not. Finally, a bidirectional LSTM layer does not change the accuracy and precision of the unidirectional layer, but it does moderately lower the perplexities (and $\gamma_{1}, \gamma_{3}$ which we discuss next).

Fortunately, the model learns to avoid predicting unphysical segments. This is evidenced by $\gamma_1$ being close to its optimal value of $1$. While $\gamma_3 \approx 2$ is further from the optimal values of $3$, in practice the rate illegal jumps (a transition from a legal segment to an illegal one) are observed at is low, as recorded in Table~\ref{tab:illegalHops}.

\begin{table}[htbp]
    \centering
        \caption{\textbf{Results}. The NYC results are based on taxi trajectories occurring in Manhttan on 01/18/11. There were $\approx 10^5$ taxi trips on this day, from which we subsampled $N = 5 \times 10^3$ which corresponded to $1.3 \times 10^5$ sequences (larger amounts of data gave memory errors). The results from the Shanghai sub-cities are based on data from 04/02/15 and since the total number of trips was $\approx 10^4$, no subsampling was necessary. The Michigan data was taken from personal car data collected in 2014, as described in Appendix~\ref{appendix:ds}. All the trips were used which amounted to $10^4$ sequences (an order of magnitude smaller than the other datasets). The figures quoted for the perplexities are the mean perplexity $\langle perp( \mathbf{x}_i \rangle_i$  of each sequence $\mathbf{x_i}$ where $perp(\mathbf{x_i}) = \prod_{i=n}^n ( prob(S_{i},S_{i+1},S_{i+n}))^{-1/n}$. The NYC and Michigan data contained a small number (between 1 and 3) of outlier perplexities $ perp(x_i) = \infty$ stemming from out-of-sample issues ($\mathbf{x}_i$ in the test set not present in the training set). We removed these outliers when calculating the mean. Note we have rounded all figures to two decimal places.}
    \begin{tabular}{lrrrrr}
    \toprule
         Dataset: uni (bi)-directional & Accuracy & Precision & Perplexity & $\gamma_1$ & $\gamma_3$ \\
         \midrule
         NYC w/o features  & 0.91 (0.91) & 0.79 (0.79)  & 2.51 (2.12) &  0.98 (1.00) & 1.77 (2.03) \\
         NYC w/ features  & 0.91 (0.91) & 0.80 (0.80)  & 2.51 (2.14) & 0.98 (1.00) & 1.75 (1.99) \\
         Yangpu  &  0.83 (0.82) & 0.73 (0.74) & 1.85 (1.57) & 1.00 (1.00) & 2.11 (2.37) \\
         Changning  & 0.82 (0.83) & 0.74 (0.75) & 1.78 (1.52) & 0.98 (0.98) & 2.03 (2.22)  \\
         Hongkou  & 0.85 (0.85) & 0.77 (0.78) & 1.73 (1.48) & 0.99 (0.99) & 1.89 (2.24)  \\
         Jingan  & 0.81 (0.81) & 0.74 (0.74) & 1.99 (1.59) & 0.98 (0.98) & 1.89 (2.13) \\
         Putuo  & 0.82 (0.82) & 0.72 (0.73) & 1.93 (1.61) & 0.99 (0.99) & 1.91 (2.12) \\
         Xuhui  & 0.84 (0.84) & 0.75 (0.76) & 1.71 (1.49) & 0.98 (0.99) & 1.96 (2.16) \\
         Michigan  & 0.58 (0.58) & 0.65 (0.65) & 14.4 (13.8) & (0.91) (0.92) & 1.05 1.13) \\
         \bottomrule
    \end{tabular}
    \label{tab:classification}
\end{table}

\vspace{2 cm}

\begin{table}[htbp]
    \centering
        \caption{\textbf{Illegal jumps}. The illegal jump rate $s$ was calculated as the fraction of illegal jumps in generated trajectories of length $L = 500$. Trajectories are generated greedily: by picking the segment / node with highest probability returned by the model. Averages and other quantities are based on ensembles of size $10^3$. Most cities have small rates of illegal hops. Notice that though the standard deviation of all cities is small, the max $s$ is high, meaning that occasionally an outlier is generated.}
    \begin{tabular}{lrrr}
    \toprule
         Dataset & Mean $s$ & Std $s$ & Max $s$ \\
         \midrule
         NYC  & 0.01  & 0.03 & 0.22 \\
         Yangpu  & 0.01  & 0.04  & 0.8  \\
         Changning  & 0.04 & 0.001 & 0.06  \\
         Hongkou  & 0.02  & 0.02 & 0.37  \\
         Jingan  & 0.01  & 0.03  & 0.25  \\
         Putuo  & 0.10 & 0.09 & 0.23 \\
         Xuhui  & 0.04 & 0.002 & 0.05  \\
         Michigan  & 0.22 & 0.12 & 0.4  \\
         \bottomrule
    \end{tabular}
    \label{tab:illegalHops}
\end{table}

\begin{figure*}[!htbp]
    \centering
    \includegraphics[width=0.375 \linewidth]{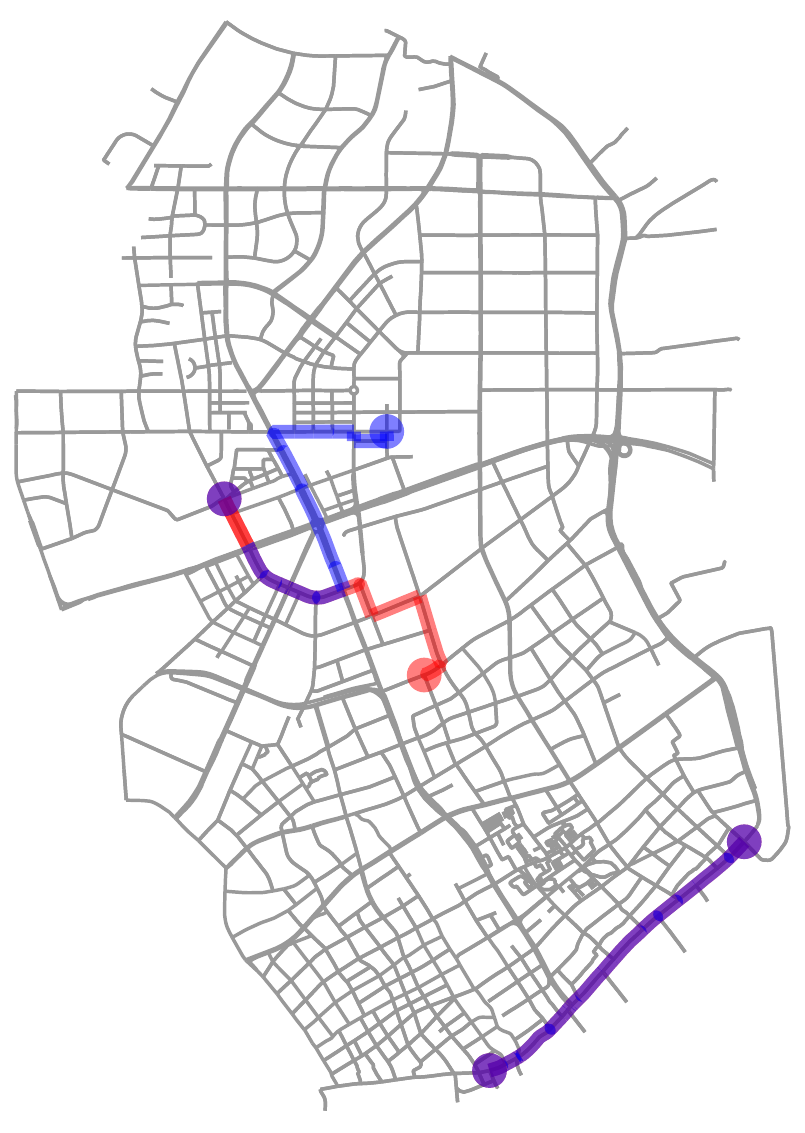}
    \caption{Yangpu street network. Red lines show empirical trajectories, blue lines show predicted trajectories. One prediction matches the data exactly (bottom) while the other prediction diverges after a period (middle). Recall this does not mean the model is displaying erroneous behavior (because taking different turns at intersections is not invalid behavior).}
    \label{fig:street-network}
\end{figure*}

We next explored if a fleet of vehicles modeled with the trained RNN produced realistic mobility patterns. We characterized mobility patterns with the street popularities $p_i$, defined as the relative number of times street segment $i$ was traversed by the vehicles during a given reference period. For each dataset we computed both the empirical $p_i$ along with the `model' $p_i$, those produced by a fleet of RNN-taxis. The latter were found by generating $N_T$ trajectories of length $\langle L \rangle$ where the empirical $N_T$ and $\langle L \rangle$ were used (i.e. we calculated $N_T$ and $\langle L \rangle$, the mean trip length, from the datasets). Trajectories were generated by feeding random initial locations $\mathbf{x}$ and greedily sampling from the RNN (recall the RNN produces a probability $z_i$ for each street $i$; so by greedily we mean we take the max of these $z_i$. We also performed experiments where streets were sampled non-greedily, w.p. $z_i$ but found no significant differences in the results; see Figure~\ref{fig:ps-prob-model.pdf}). The initial conditions $\mathbf{x}$ (we recall is a sequence of $k$ segments) were found by choosing an initial node uniformly at random, then choosing a neighbour of this node again at random, and repeating until $k$ segments were selected. In Figure~\ref{fig:street-network} we show some empirical and generated trajectories on the Yangpu street network.

\begin{figure*}[!htbp]
    \centering
    \includegraphics[width = \linewidth]{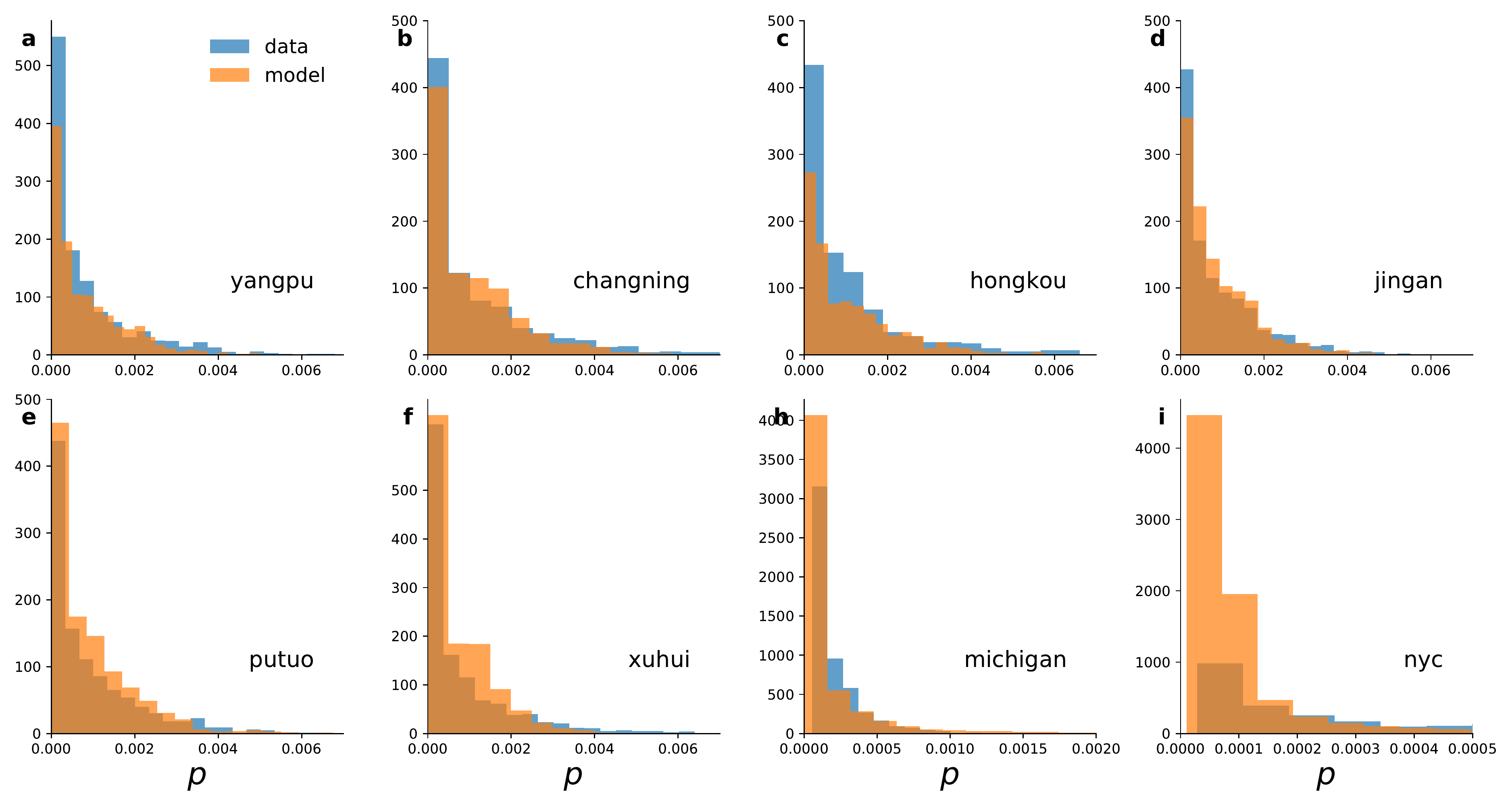}
    \caption{Histograms of street popularities calculated from empirical trajectories and generated (i.e. from the RNN model) data for each. The popularity $p_i$ of a given node is defined as the relative number of times each node is traversed by a vehicle in a given reference period $\mathcal{T}$ (we assume $\mathcal{T}$ is large enough so that $p_i$ are approximately stationary). As can be seen, the model $p_i$ well capture the empirical $p_i$. The model $p_i$ were found by simulating $N_T$ trips of length $L$, where $N_T$ was the total number of trips occurring in the dataset for each city, and $L$ as the average length of a trajectory. That way, model and empirical $p_i$ were compared evenly. To quantiatively compare the distributions we used the KS-statistic $D$ (for which smaller values indicate better fits). The values found were $D = 0.25, 0.26, 0.18, 0.21, 0.21, 0.27, 0.64, 0.42$ in panel order.}
    \label{fig:ps-histogram}
\end{figure*}

Figures~\ref{fig:ps-histogram} and~\ref{fig:ps-pdf} show the distribution of model $p_i$'s well approximate the empirical distribution of $p_i$'s for all datasets. This is an encouraging finding, although other models in the literature can also reproduce realistic segment distributions. For example, the `taxi-drive' model of taxis studied in \cite{o2019quantifying, o2019urban} also produces realistic distributions of $p_i$. But the \textit{spatial} distributions of the taxi-drive $p_i$ (as opposed to raw `counts') are less accurate. We were curious if the RNN could improve upon this situation so in Table~\ref{tab:correlation} we compare the models by computing the pearson correlation coefficient $r$ between the empirical $p_i$ and the model $p_i$ (we only perform this part of analysis for the taxi data, so ignore the Michigan personal-car data). We also show the $r$ for a baseline bigram model where the Markovian transition probabilities are used $P(S_i | S_{i-1})$ and are simply estimated from raw frequencies.

\begin{figure*}[!htbp]
    \centering
    \includegraphics[width=0.90 \linewidth]{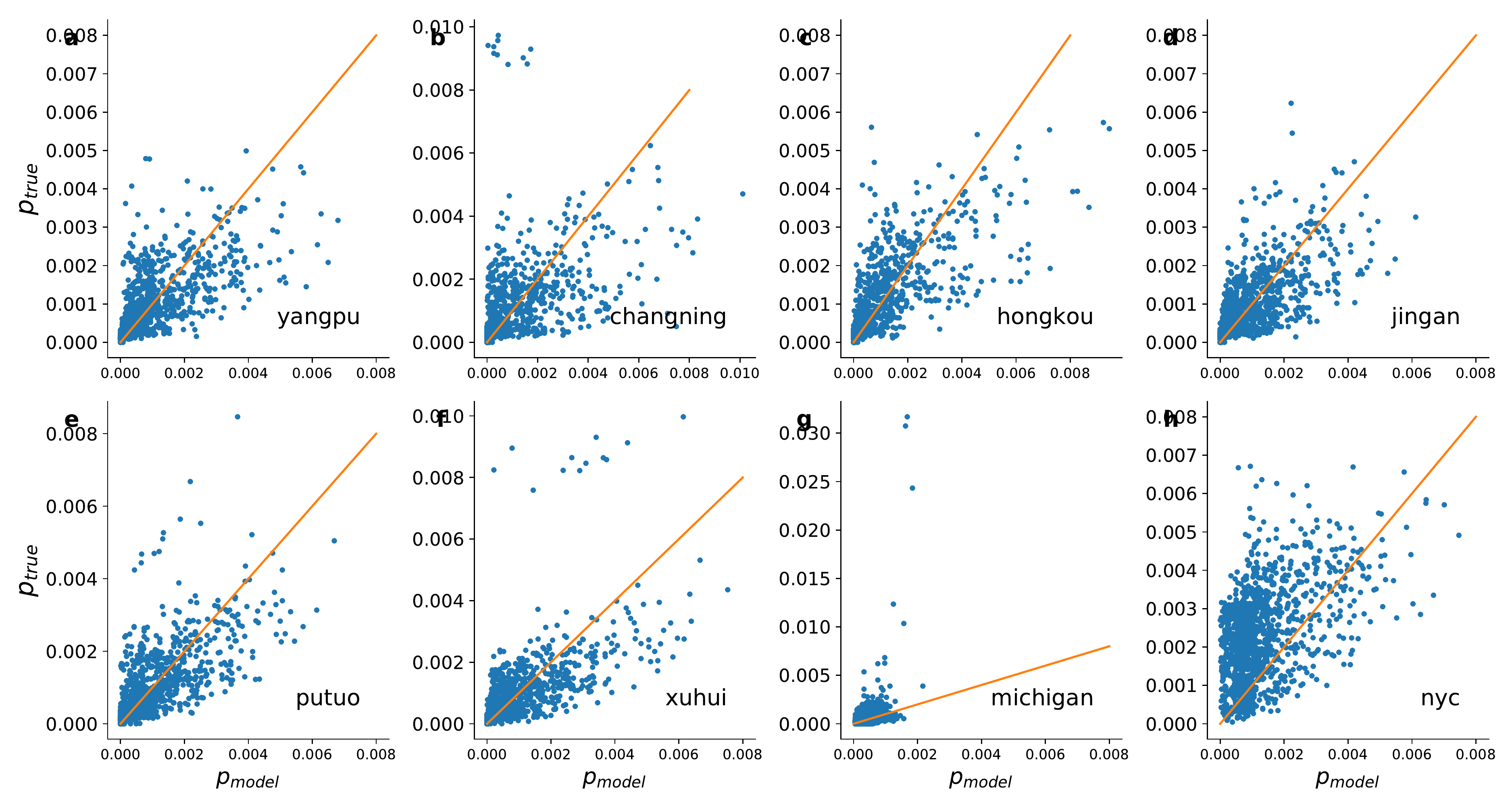}
    \caption{Scatter plots of model $p_i$ and empirical $p_i$. The pearson correlation values in panel order are $r = 0.71, 0.46, 0.728,0.687,0.705,0.666, 0.46, 0.48$.}
    \label{fig:ps-correlate}
\end{figure*}

As seen the RNN improves upon the other models, having the largest $r$ values for all datasets. In Figure~\ref{fig:ps-correlate} we give a visualization of this correlation by showing a scatter plot of $p_{i,model}$ and $p_{i,data}$ for the RNN model. 


 \begin{table}[htbp] 
 \centering 
 \caption{\textbf{Correlation coefficients}. Pearson correlation coefficients between model $p_i$ and empirical $p_i$ for different models of taxis. The taxi-drive was run for $10^5$ time units after which the distribution of the $p_i$ were approximately stationary. The $\beta$ parameter in the taxi-drive model were $\beta = 1.5, 2.75, 0.75, 0.75, 2.0, 0.5$. These were chosen by strobing over a range of $\beta$ values and choosing the value which best matches empirical data as measured by the KS statistic.} 
\begin{tabular}{lrrrr} 
 \toprule 
Dataset & Taxi-drive & Bigram & RNN \\ 
 \midrule 
 NYC & 0.51 & 0.29 & 0.69 \\ 
 Yangpu & 0.48 & 0.40 & 0.71 \\ 
 Changning & 0.38 & 0.37 & 0.46 \\ 
 Hongkou & 0.55 & 0.27 & 0.73 \\ 
 Jingan & 0.64 & 0.22 & 0.69 \\ 
 Putuo & 0.40 & 0.19 & 0.70 \\ 
 Xuhui & 0.60 & 0.23 & 0.67 \\ 
\bottomrule 
 \end{tabular} 
 \label{tab:correlation} 
 \end{table}

\section*{Discussion}
Our primary result is that RNN's can well capture the mobility patterns of real taxis, as reflected in street popularity distributions $p_i$, And to a lesser extent, capture the mobility patterns of personal cars (the Michigan dataset showing lower performance). In particular when it comes to spatial distributions of these $p_i$, the RNN's improve over current models. This finding could be useful for mobile sensing applications, where spatially accurate $p_i$ could allow fine-tuned sensing \cite{o2019quantifying}. We also found that, surprisingly, using extra features (GPS and time of trip) did not improve model performance. The same is true of using bi-directional LSTM's over uni-directional ones; only a small decrease in perplexity was observed. 

Broadly speaking, our results validate the use of machine learning techniques in modeling mobility systems, a goal which, given the loom of autonomous vehicles and other forms of smart mobility, seems timely. Future work could investigate how well RNN's model other mobility patterns, such as those of private cars, or even other entities like bacteria and cells.

\section*{Acknowledgements}
 We thank Allianz, Amsterdam Institute for Advanced
Metropolitan Solutions, Brose, Cisco, Ericsson, Fraunhofer Institute, Liberty
Mutual Institute, Kuwait–MIT Center for Natural Resources and the Environment, Shenzhen, Singapore–MIT Alliance for Research and Technology,
UBER, Victoria State Government, Volkswagen Group America, and all of
the members of the MIT Senseable City Laboratory Consortium for supporting this research.


\bibliographystyle{unsrtnat} 








\appendix


\section{Datasets}
\label{appendix:ds}

The taxi dataset has been obtained from the New York Taxi and Limousine Commission for the year 2011 via a Freedom of Information Act request, and is the same dataset as used in previous studies \cite{santi2014quantifying}. The dataset consists of a set of taxi trips occurring between 12/31/10 and 12/31/11. There are $N_{trips} = O(10^5)$ trips per day. Each trip $i$ is represented by a GPS coordinate of pickup location $O_i$ and dropoff location $D_i$ (as well as the pickup times and dropoff times). We snap these GPS coordinates to the nearest street segments using OpenStreetMap. We do not however have details on the trajectory of each taxi -- that is, on the intermediary path taken by the taxi when bringing the passenger from $O_i$ to $D_i$. As was done in \cite{santi2014quantifying}, we approximate trajectories by generating 24 travel time matrices, one for each hour of the day. An element of the matrix $(i,j)$ contains the travel time from intersection $i$ to intersection $j$. Given these matrices, for a particular starting time of the trip, we pick the right matrix for travel time estimation, and compute the shortest time route between origin and destination; that gives an estimation of the trajectory taken for the trip. 

The Shanghai datasets were provided by the `'1st Shanghai Open Data Apps 2015" (an annual competition) and contain the full GPS-trace of a taxi. The span the week 03/01/14 – 03/07/1. There are $N_{trips} = O(10^5)$ trips per day.We matched trajectories to OpenStreetMap (driving networks) following the idea proposed in \cite{newson2009hidden}, which using a Hidden Markov Model to find the most likely road path given a sequence of GPS points. The HMM algorithm overcomes the potential mistakes raised by nearest road matching, and is more robust when GPS points are sparse.

The Safety Pilot Model Deployment (SPMD) offers detailed instantaneous driving data generated by connected vehicles in real-world environment. This pilot, sponsored by US-DOT, is currently on-going in Ann Arbor, Michigan, and intends to display vehicle-to-vehicle (V2V) and vehicle-to-infrastructure (V2I) communication systems in real-life environment. The Michigan data was collected using the Safety Pilot Model Deployment (SPMD) technolgoy offers detailed instantaneous driving data generated by connected vehicles in real-world environment. This pilot, sponsored by US-DOT, is currently on-going in Ann Arbor, Michigan, and intends to display vehicle-to-vehicle (V2V) and vehicle-to-infrastructure (V2I) communication systems in real-life environment. The data is feature rich -- containing data on many aspects of driving behavior -- but in this work we are just interested in GPS coordinates,  which we match to street segments using the same method as above. The dataset contained $N_{trips}$ = 2940 trips occuring in 2014. Note this is substantially smaller than the other datasets.

Note that in both of the taxi datasets, we do not have any information on the taxis movements when it is empty, looking for passengers. 


\section{Metrics}
\label{appendix:metrics}
We here formally define the metrics used in the paper: (i) accuracy (ii) precision and (iii) perplexity. Let $ \hat{y}_i = f(\bf{x_i}) $ be the $i$-th predicted segment and $y_i$ be the true in the sequence $\mathbf{x}_i$. Then the accuracy is the fraction of correct predictions
\begin{equation}
    accuracy = (N)^{-1} \sum_i^N 1 (\hat{y}_i = y_i)
\end{equation}
\noindent
where $N$ is the number of samples in the test set and $1$ is the indicator function. The precision for a given class is the fraction of predictions for that class that were correct:
\begin{equation}
    precision_A = \frac{\sum_i 1 (\hat{y}_i = A, y_i = A)}{\sum_i 1 (\hat{y}_i = A)}
\end{equation}
\noindent
In a multi-class setting (which we have in this work) the average precision over all the classes is reported. The perplexity is the probability of generating the test set. Recall the test set has form $(\mathbf{x}_i, y_i)$ where $\mathbf{x}_i  = (S_{i-k},S_{i-k+1},..S_{i-1})$ and $y_i = S_i$. Concatenate these $w_i = (S_{i-k}, \dots, S_i)$. Then the perplexity for $w_i$
\begin{equation}
    perplexity(w_i) = Prob(w_{i-k}, \dots, w_{i})^{-1/(k+1)}
\end{equation}
The mean perplexity over the test set, $perplexity = \langle perplexity(w_i) \rangle_i$ is then taken (and reported in the main text).


\section{Hyperparameter optimization}
\label{appendix:ho}

We optimized the following hyperparameters: learning rate over (1e-1, 1e-4, 1e-8), embedding dimension over (16,32,64,128,256), hidden dimension over (16,32,64,128,256), epochs over (10,20,40), and batch size over (32,64). We used python's RandomizedSearchCV which picks $N_{iter}$ points at random in the grid in hyperparameter space. We picked $N_{iter} = 10$ and used $3-$ fold validation. The network architecture consisted of one embedding layer, two LSTM layers followed by two Dense layers. The optimal hyperparameters chosen were learning rate = 1e-4, hidden dimension = 64, epochs = 40, embedding dimension 256.

As discussed in the main text, we tested for the optimal sequence length separately. Figure~\ref{fig:seqLength}(a) shows how the accuracy and precision for the NYC dataset vary with sequence lengths. As can be seen, the accuracy and prediction saturate at seq length $=3$, so we used that value throughout the paper.

In all experiments in the main paper, the hyperparmeters outlined above were used, and a sequence length of 3 was used. Note that though the above hyperparameters were found by training on the NYC data only, we assumed they are optimal for the Shanghai datasets also. The hyperparameters for the Michigan datasets were: lr = 1e-3, hidden dimension = 64, epochs = 40, and embedding dimension = 256.


\section{Supplementary Figures}
\label{appendix:sf}

\begin{figure*}[!htbp]
    \centering
    \includegraphics[width=0.75 \linewidth]{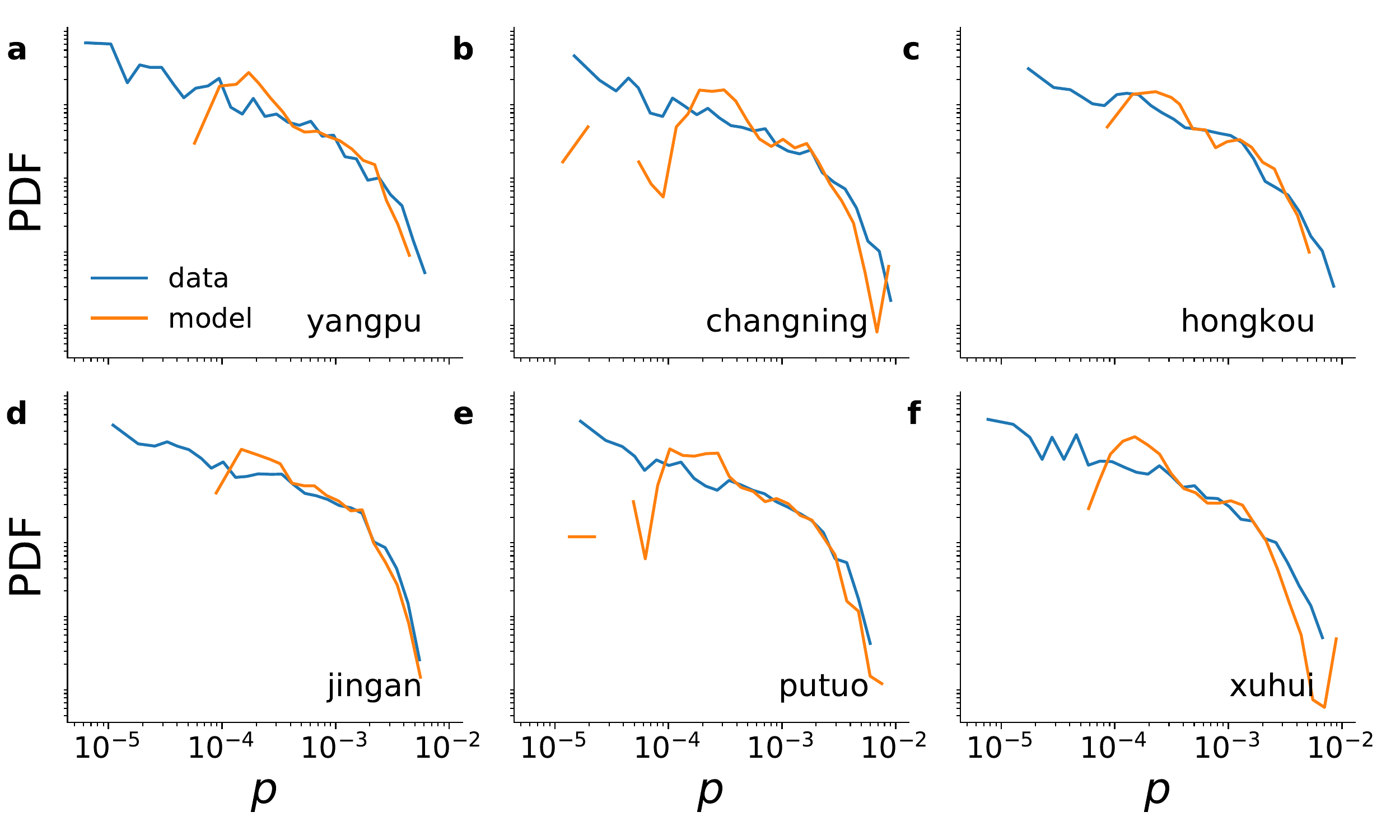}
    \caption{Counterpart to Figure~\ref{fig:ps-histogram} where we show the probability density function PDF of the node popularities instead of
    histograms. Note the good agreement between the model and the data.}
    \label{fig:ps-pdf}
\end{figure*}

\begin{figure*}[!htbp]
    \centering
    \includegraphics[width=0.5 \linewidth]{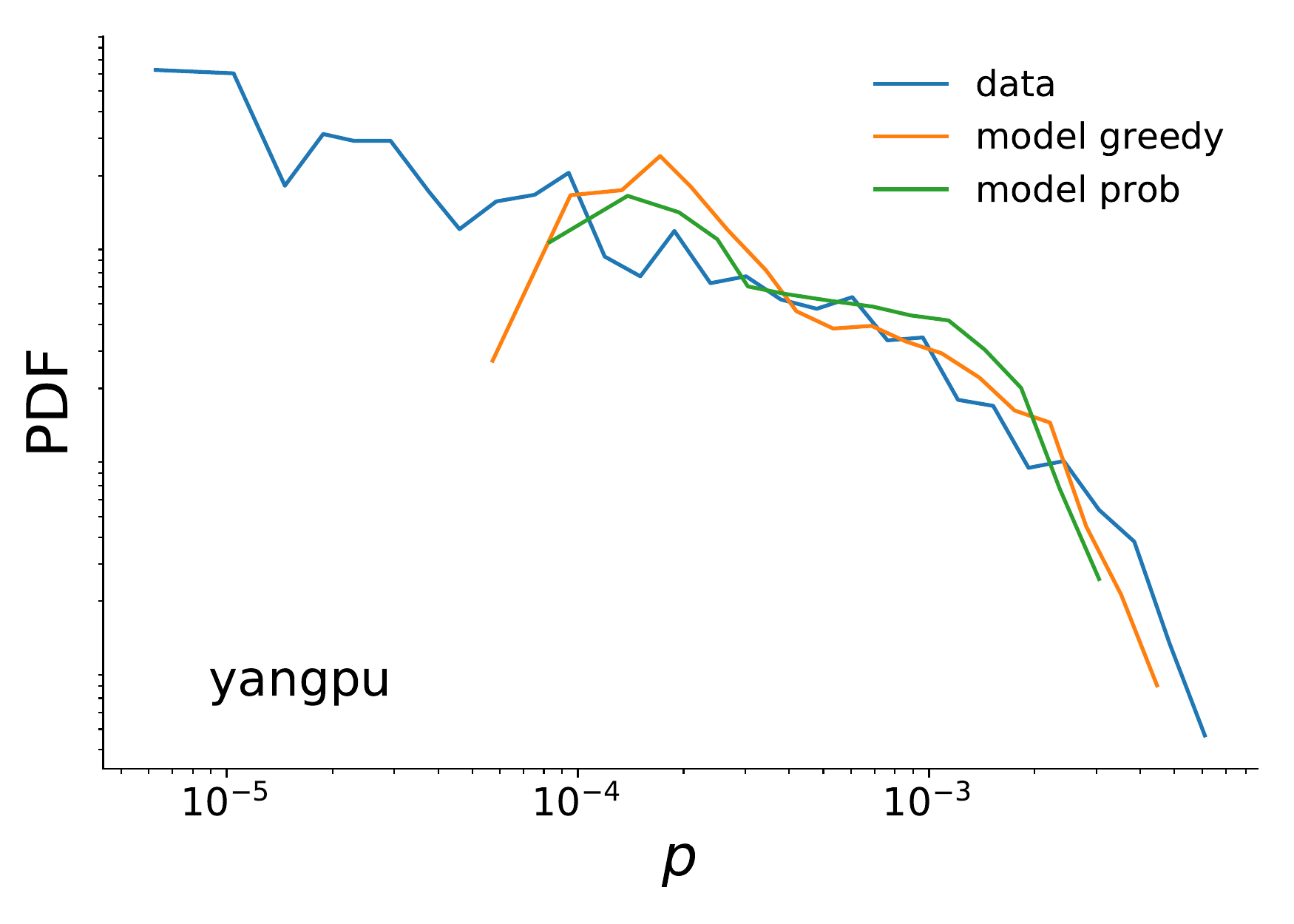}
    \caption{Probability density functions of node popularities $p_i$ from empirical data, the greedy model, and probabilistic model. Both the greedy $p_i$ and probabilistic $p_i$ match the data well.}
    \label{fig:ps-prob-model.pdf}
\end{figure*}

\begin{figure*}[!htbp]
    \centering
    \includegraphics[width=0.75 \linewidth]{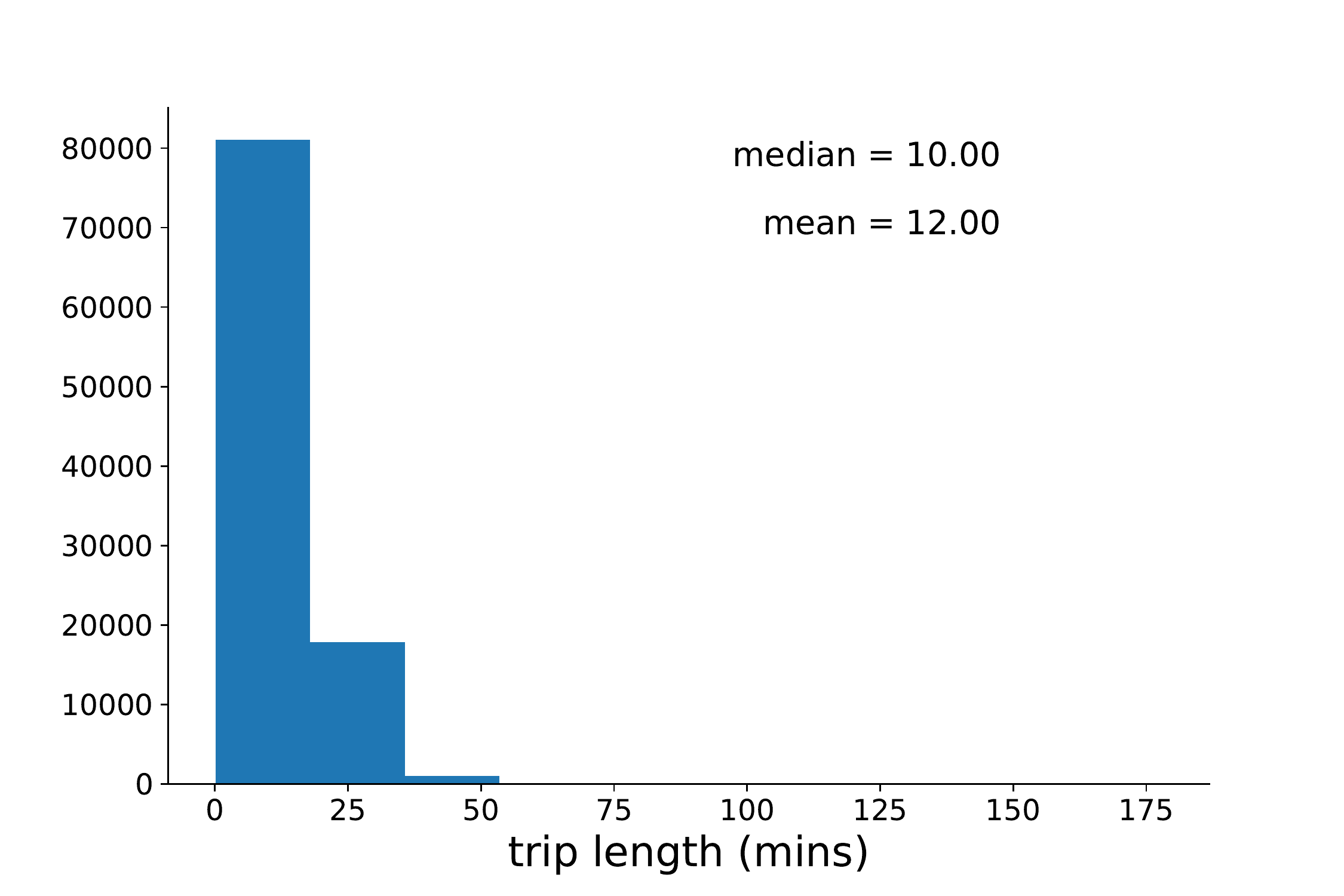}
    \caption{Distribution of taxi trip lengths in NYC on 1/18/19.}
    \label{fig:trip-length}
\end{figure*}

\end{document}